\documentstyle[12pt]{article}

\begin{document}

\font\fortssbx=cmssbx10 scaled \magstep2

%
%
%

%
\medskip

\begin{center}
{\Large\bf\boldmath Military use of
depleted uranium: assessment of prolonged population exposure}
\rm
\vskip1pc
\medskip
{\Large  C. Giannardi $^{a}$ and D. Dominici$^{b,c}$}
\end{center}
\begin{center}
\vspace{5mm}
{\it{$^a$  Fisica Ambientale, Dipartimento di Firenze, ARPAT
\\
c.giannardi@arpat.toscana.it\\
$^b$Dipartimento di Fisica, Universit\`a di Firenze
\\
$^c$I.N.F.N., Sezione di Firenze\\
dominici@fi.infn.it 
\\
}}
\end{center}
\bigskip
\begin{abstract}
This work is an exposure assessment for a population 
 living in an area contaminated by use of depleted uranium (DU) weapons. 
RESRAD 5.91 code is used to evaluate the average effective dose delivered 
from 1, 10, 20 $cm$ depths of contaminated soil, in a residential 
farmer scenario. Critical pathway and group are 
identified in soil inhalation or ingestion and children 
playing with the soil, respectively. From available information 
on DU released on targeted sites, both critical and average exposure 
can leave to toxicological hazards; annual dose limit for population 
can be exceeded on short-term period (years) for soil inhalation.
 As a consequence, in targeted sites cleaning up 
must be planned on the basis of measured concentration, when available, 
while special cautions have to be adopted altogether to reduce unaware 
exposures, taking into account the amount of the avertable dose.
\noindent
\end{abstract}
\newpage

\section{Introduction}

Munitions containing depleted uranium (DU) have been used by
NATO and US forces during the war operations in Iraq (1991), Bosnia
(1994), Kosovo and Serbia (1999).
Recently some information on 112 sites targeted by DU weapons
 in Kosovo has been supplied
by NATO to the United Nation Environmental
Program Balkans Task Force (UNEP BTF);  on November 2000 measurements 
to detect contamination have been undertaken by a UNEP team in 11 
among the 112 sites.

Aim of this paper is outlining some aspects
of the  exposure of people living  in an area contaminated 
by DU, on the basis of official available information and of simulations, 
looking for main pathways of average and critical exposure.

Individuation of pathways of high exposure could allow to advice to 
population; average dose assessment, together with measures of DU
 concentration 
in soil, will make  delimitation of areas to be cleaned up possible.

\section{Military use of depleted uranium}

The Gulf war against Iraq in 1991 was the first one known
where DU rounds have been 
used in large quantity (approximately 300 tonnes)  \cite{4,5}. 
The consequences on the health of the Iraqi population and 
of the US veterans are still under study.
DU exposure at the moment is not considered the most probable 
 cause of the Gulf War Syndrome experienced 
by hundreds thousand veterans \cite{qs}; on the other hand, 
the effects of the DU left over the Iraqi territory are difficult to 
show, due to the large number of toxic substances dispersed in 
the environment during the war and the deterioration of the sanitary 
situation caused by the {\it embargo} to which the country is 
submitted from 1991 (cited work in \cite{8}, app.3).

Reports on potential effects on human health and environment
from the use of DU have appeared during the last years:
studies on risk assessment for the Jefferson Proving Ground,
a US facility for testing DU munitions,
have been performed \cite{ebi98}; the risk for population 
for the Kosovo conflict
and for the Gulf war has been also considered \cite{8,24}.

DU can  be obtained as by-product in the enrichment 
process of natural uranium for 
the production of nuclear fuel and for military applications; 
as the ore extracted natural uranium, DU 
 is associated to a reduced chain of radioactive isotopes, 
formed by $^{238}U$ and $^{235}U$ decay products having shorter decay times: 
$^{234}Th$ (24 days), $^{234m}Pa$ (1.17 min) and $^{234}Pa$ (6.7 hours), 
$^{231}$  Th(25.5 hours). DU can also be obtained by the 
reprocessing of nuclear power plant spent fuel, and so traces 
of transuranic elements and $^{236}U$ can be present. 
According to official information, DU used by the U.S. Department 
of Defence contains approximately 0.2$\%$ of $^{235}U$ 
and traces of $^{234}U$, $^{236}U$. 
Following the indications in  \cite{AEPI95,RAND}, we  will assume 
the uranium isotopic composition of DU given in Table \ref{tab8}. 
\begin{table}[htb]
\caption
{\it    Assumed depleted uranium composition. $A_i$ is the 
specific isotopic activity, $A_{DU}$ is the activity concentration 
per $mg$ of DU.
} 

\begin{center}
\begin{tabular}{|c|c|c|c|c|}
\hline 
& \% &$ T_{1/2}$
 &$A_i$&$A_{DU}$  \\
&  &$( y)$
 & $(Bq/mg)$ &$(Bq/mg)$\\
\hline
$^{238}$U &99.796 &4.5 10$^9$
&12.4&12.375\\
\hline
$^{235}$U &0.2 &0.7 10$^9$
&80&0.160\\
\hline
$^{234}$U &0.001 &2.5 10$^5$
&2.3 10$^5$&2.300\\
\hline
$^{236}$U &0.003 &2.3 10$^7$ 
&2.4 10$^3$&0.072\\
\hline
$\sum_{U}$ &100&
& &14.907\\
\hline
\end{tabular}
\end{center}
\label{tab8}
\end{table}
DU specific  activity is in part due to uranium isotopes 
(14.9 $Bq/mg$, 36\%), and for the residual part to beta emitting short-life 
decay products (64\%); among the transuranic elements
official 
information is available only for $^{239}Pu$ (2.4 10$^4$ years), whose content 
is estimated in 11 ppb \cite{10}. DU specific  activity
is not substantially affected by the declared amount of traces elements. 

Metallic uranium has a high density (19 $g/cm^3$), is pyrophoric and 
cheaper than  tungsten, and so has been attractive for U.S. Army for the 
production of armor piercing ammunition since 1960s. 
Tungsten alloys have been preferred up 1973, when a DU alloy 
with 0.75$\%$ of titanium (U-3/4Ti) was adopted for ammunition 
made by a thin cylinder in DU alloy encased with lighter material. 
Systems of DU  weapons are owned or under development in different 
countries (Saudi Arabia, France, United Kingdom, Israel, Pakistan, 
Russia, Thailand and Turkey) \cite{RAND}. 

Use of DU ammunition causes exposure of people soon and after, because DU is 
dispersed as aerosol when the projectile strikes a hard target and then falls 
out on a limited area \cite{AEPI95}. Contamination of all environmental 
matrices takes place and health effects on people living nearby must be 
taken in account, both for toxicological damage and for radiological risk. 
Among different isotopes present in DU as declared, $^{238}U$,$^{234}U$ 
and $^{235}U$ are of concern in risk assessment. 
For chemical hazard, kidney is identified as the target organ, 
whatever the path of assumption  \cite{13}. 
Due to prevalent  short-range emitted radiation, the risk associated 
with exposure to ionizing radiation mainly derives from ingestion and 
inhalation of radioactive material; external 
irradiation from soil is less relevant.

\section{Dispersion of DU in the environment and exposure
of the population}

DU contained in projectiles, spread out as aerosol
in air after striking the target, 
falls out producing environmental and food chain contamination. 
Possible occurring of chemical hazard and entity of radiation dose must be 
assessed for  people living in the area, 
taking into account both average and critical group exposure.

DU concentration in the soil is the starting point; while waiting for 
measurements of contamination in Iraq, Bosnia,  Kosovo
and Serbia,
we present computed radiation doses and associated
concentrations for 
different contaminated soil thickness, as soil mixing will 
extend the initial superficial deposition to underlying layers in not 
undisturbed areas.
Available soil measured DU concentrations in contaminated 
sites that we are aware  of, are the following:
\begin{itemize}
\item{at Jefferson Proving Ground area an average 
$\sum U$ concentration of 318 $Bq/kg$  was reported \cite{26};
more recently a lower and an upper bound of the concentration
ranging from 592  $Bq/kg$ to 13690 $Bq/kg$ was also
measured \cite{ebi98};}
\item{among  the areas where the US personnel lived in the Gulf
region (outside Iraq) the highest DU concentration (433 $Bq/kg$)
was measured in the Iraqi Tank Yard 
(the area where captured Iraqi 
equipment is stored in Kuwait) \cite{dod2000};}
\item{in some sample analyzed by the RFY scientists a specific activity 
of $^{238}U$ up to  $2.35 ~10^5~Bq/kg$  was detected \cite{RFY}.}
\end{itemize}
Following the hypothesis assumed in the BTF report, 
we have assumed as a reference value a DU contamination 
of $1000 ~Bq/kg$   of soil over an area of 
$A=10000~ m^2$, in the hypothesis of 10 $kg$ of DU entirely 
dispersed in the impact as aerosol of uranium oxides,
 contaminating 1 $cm$ of soil. 
With the composition given in Table \ref{tab8} initial 
activities per $kg$ of soil for $^{238}U$,  $^{235}U$, $^{234}U$
and $^{236}U$ are respectively
  830 $Bq$, 
11 $Bq$, 154 $Bq$ and 5  $Bq$. 

Average effective dose is conservatively assessed using the 
residential farmer scenario. The following pathways are considered: 
external irradiation from soil, inhalation from resuspended dust, 
ingestion of contaminated soil and water, ingestion of plants and 
animal products grown in site and ingestion of fish grown in 
a pond contaminated by groundwater. Different pathways are considered 
for plant contamination due to first root uptake (water independent) 
and due to secondary root 
uptake from use of contaminated water (water dependent). 
Radon inhalation is excluded. 
RESRAD 5.91 \cite{resrad} code is used, all parameters default 
except for the ones given in Table \ref{resrad}. Estimates of dose 
to individuals and population for risk in contaminated sites have been
performed  by EPA employing primarily the code RESRAD (for related 
work see \cite{HP1,HP2}).

 RESRAD default libraries values have been corrected 
to give effective dose \cite{72} rather than 
equivalent effective dose \cite{icrp26}: 
due to the algorithm used by RESRAD, 
anyway, values for external irradiation $E_G$ 
in  Tables  \ref{tab2} and \ref{tab3} have been impossible to modify, 
and are approximate by 10\% maximum defect. 


In Tables \ref{tab1}, \ref{tab2},  \ref{tab3} and \ref{conc}
we show average 
annual effective doses and
corresponding DU concentrations
in water and vegetables for three different soil thickness, respectively 
$1~cm$,  $10~cm$ and $20~cm$. The following
quantities
 are given at different times, 
from the first year to about two  hundred  years after maximum dose, 
for main pathways: the total dose ($E_{tot}$),  the dose from external 
irradiation from the soil ($E_G$),  from inhalation of contaminated 
dust ($E_I$), from consumption of edible plants (water independent
$E_P$, water dependent $E_P^w$) and  of water ($E_{H_2O}$).


\begin{table}[htb]
\caption
{\it RESRAD parameters different from the default value.  
} 
\begin{center}
\begin{tabular}{|c|c|c|c|c|}
\hline 
& 
this paper& 
RESRAD def\\
\hline
indoor time fraction &0.6&0.5\\
outdoor time fraction&0.2  &0.25\\ 
exposure duration& 50 $years$&30 $years$\\     
well pump intake depth& 3 $m$ &10 $m$\\ 
drinking water intake& 730 $l/y$ & 510 $l/y$\\ 
\hline
\end{tabular}
\end{center}
\label{resrad}
\end{table}


\begin{table}[htb]
\caption{\it Effective doses ($\mu Sv$) for
contaminated soil thickness 
$1~ cm$. $E_{tot}$, $E_G$, $E_I$,
$E_P$, $E_{H_2O}$, $E_P^w$
are the total dose, the ground, inhalation,
plant (water independent), water, plant
(water independent) doses.
 The initial contamination is assumed of   $1000 ~Bq/kg$    over an
area of  $A=10000~ m^2$. The symbol - means doses less than 1 $\mu Sv$.
All not specified parameters as in Table \ref{resrad}.} 
\begin{center}
\hspace{-0.6cm}
\begin{tabular}{|c|c|c|c|c|c|c|c|}
\hline
$t(y)$& 
$E_{max}  $& 
$E_G$&$E_I $&$E_P $   &$E_{H_2O}  $
&$E_P^w $\\
\hline
0&4 &4 & - & - &-&-\\
1& 3&3 &- &-&-&-\\    
3&- &- & - &-&-&-\\
300&- &- & - &-&-&-\\ 
485&4 &- &-  &-&4&-\\ 
500&4 &- &-  &-&4&-\\ 
700 &- & - &-&-&-&-\\     
\hline
\end{tabular}
\end{center}
\label{tab1}
\end{table}

The dependence of $t_{max}$ and $E_{max}$ on some hydrogeological parameters, 
mainly affecting the water dependent pathways, is shown in Table \ref{tab4}
and \ref{tab5}.  The maximum value of the dose is not much affected by 
most of the parameters considered in Table \ref{tab5} except  $K_d$.
This parameter is defined as the ratio of the mass of solute species 
observed  in the solids per unit of dry mass of the soil to the solute 
concentration in the liquids. A wide  range has been observed for uranium  
$K_d$  values \cite{shepp}. For largest value of $K_d$ the DU is retained
in surface and does not reach at least within the first 1000
years  the watertable. A measurement of the local value of this
parameter is therefore necessary to reduce the uncertainty
on the dose assessment. 

Strong dependence of maximum  inhalation dose has been found, as 
expected,  on the dust loading parameter, as shown in Table \ref{tab6}.

\begin{table}[htb]
\caption{\it Effective doses ($\mu Sv$) for
contaminated soil thickness 
$10~ cm$. All not specified parameters as in Table \ref{resrad}.
} 

\begin{center}
\begin{tabular}{|c|c|c|c|c|c|c|c|}
\hline
$t(y)$& 
$E_{max}  $& 
$E_G$&$E_I $&$E_P $   &$E_{H_2O}  $
&$E_P^w $\\
\hline
0&18 &15 & - &1&-&-\\
1&17 &14 & - &1&-&-\\    
3& 15&12 & - &1&-&-\\
10& 9&8 &-  &1&-&-\\
30& 2&2& - &-&-&-\\             
100&-&- & - &-&-&-\\
300&-&- & - &-&-&-\\
486&44&- & - &-&41&2\\  
500 &44 & - &-&-&41&2\\  
700 &- & - &-&-&-&-\\  
\hline
\end{tabular}
\end{center}
\label{tab2}
\end{table}

\begin{table}[htb]
\caption{\it Effective doses ($\mu Sv$) for
contaminated soil thickness 
$20~ cm$. All not specified parameters as in Table \ref{resrad}.
}

\begin{center}
\begin{tabular}{|c|c|c|c|c|c|c|c|}
\hline
$t(y)$& 
$E_{max}  $& 
$E_G$&$E_I $&$E_P $   &$E_{H_2O}  $
&$E_P^w $\\
\hline
0&24 &19 &1  &2&-&-\\
1& 23&18 & 1 &2&-&-\\    
3& 21&17 & 1 &2&-&-\\
10& 17&13& - &2&-&-\\
30& 8&7 & - &1&-&-\\             
100&  1& 1 &-&-&-&-\\ 
300&  -& - &-&-&-&-\\ 
499&  87& - &-&-&82&4\\ 
700 & 1& - &-&-&1&-\\     
\hline
\end{tabular}

\end{center}
\label{tab3}
\end{table}

\begin{table}[htb]
\caption
{\it    DU concentrations in the water $C_{H_{2}O}$
and in the edible plants (water dependent)
$C_P^w$ at the maximum dose time.
} 
\begin{center}
\begin{tabular}{|c|c|c|}
\hline 
& $C_{H_{2}O}~(Bq/l)$  &$C_P^w~(Bq/kg)$ \\
\hline
1 $cm$&1.11
&1.48\\
\hline
10 $cm$&1.15
&1.85\\
\hline
20 $cm$&2.25
&3.74\\
\hline
\end{tabular}
\end{center}
\label{conc}
\end{table}

As already outlined, presented doses and concentrations have been obtained
 from an average value of soil contamination, in order to assess the 
average exposure of population. Whatever the average value considered, 
anyway, highly inhomogeneous soil concentrations must be expected in 
the contaminated area, both for sparse aerosol deposition and for 
oxidation of DU fragments: concentrations up to 12\% in weight have been 
reported \cite{28}. In order to assess the dose to critical population group, 
this must be taken in account, especially if inhalation of soil was the 
critical pathway: inhalation of 0.1 $g$ of soil with maximum reported DU 
contamination, equal to 12 $mg$ DU, corresponds to 1.44 $mSv$; ingestion of 
1 $g$ of soil, equal to 120 $mg$ DU, corresponds to 0.08 $mSv$. 

A scenario, in which permanence in dusting air and ingestion of soil are 
possible,  is the one for children playing with soil.
From the presented dose assessment and considerations
 children playing with soil
may be identified as the critical population group,
 with 
inhalation and/or ingestion of contaminated soil as critical pathway. 
Evidently, average and critical doses are somehow competitive, because 
the higher fraction of DU is dispersed as aerosol, the lower part of it 
can rest in soil as fragment, being presence of fragments the main cause of 
hot spots in soil contamination.

It must be outlined that the amount of DU considered in the simulation 
corresponds to 37 A-10 /GAU-8 ammunitions. According to the 
available information, a much larger number of projectiles has been 
fired on each site (between 50 and 2320, average 300) and up to now 
unknown is the extension of targeted sites. Both for average and critical 
exposure, anyway, more realistic dose assessment will be possible only 
when measured contamination data will be known, scaling the values in the
tables  for the appropriate factor.

Increment of inhalation 
dose attributable to $^{239}Pu$ presence in DU is officially 
estimated in 14$\%$ \cite{10}:
with 11 ppb of $^{239}Pu$ in DU RESRAD gives a maximum dose
increment of 0.6\%.

\begin{table}[htb]
\caption
{\it Effective doses for contaminated soil thickness  
$10~ cm$, unsaturated zone thickness 3.90 $m$, for different values of
the well pump intake depth (WPID). ($C_{H_2O}$  and
$C_P^w$ are the concentrations of DU in the water and in the
plants (water dep)). All not specified parameters as in Table \ref{resrad}.}

\begin{center}
\begin{tabular}{|c|c|c|c|c|}
\hline 
WPID$(m)$& 
$t_{max}(y)$& 
$E_{max}  (\mu Sv)$& 
$C_{H_2O}  (Bq/l)$
&$C_P^w(Bq/kg)$\\
\hline
1 &398 & 103 &2.7&4.5\\
2 &417 & 65&1.7&2.8\\             
4 &564 & 33 &0.7&1.4\\             
\hline
\end{tabular}
\end{center}
\label{tab4}
\end{table}

\begin{table}[htb]
\caption
{\it Contaminated soil thickness  
$10~ cm$. All not specified parameters as in Table \ref{resrad}.
} 
\begin{center}
\begin{tabular}{|c|c|c|c|c|}
\hline 
& 
$t_{max}(y)$& 
$E_{max}  (\mu Sv)$\\
\hline
prec.rate $(0.9-1.1)m$ &$537-435 $&$43.0-44.1 $\\
watershed area $(10^6\pm 10^5)m^2$  &$486 $ &$43.6 $\\ 
well pumping rate $(200-300)m^3/y$&$486 $ &$43.6 $\\     
distrib.coeff. $K_d$ $(20-100)cm^3/g$ &$215-0$&$ 118-19$\\ 
\hline
\end{tabular}
\end{center}
\label{tab5}
\end{table}


\begin{table}[htb]
\caption
{\it 
Average dose from inhalation at $t=0$ for different values of
the dust loading parameter.
Contaminated soil thickness  
$10~ cm$. All not specified parameters as in Table \ref{resrad}.
} 
\begin{center}
\begin{tabular}{|c|c|c|c|}
\hline 
& $100~ \mu g/m^3$ &$ 1~ mg/m^3$ &$5~ mg/m^3$  \\
\hline
inhal. dose $(\mu Sv)$\ &- &3&16\\
\hline
\end{tabular}
\end{center}
\label{tab6}
\end{table}
\vskip1truecm

\section{Normative and recommendations framework}

Before discussing compliance of average assessed doses 
and exposure with international standards set to prevent 
from toxicological damage and limit ionizing radiation risk, 
we shortly line out an aspect relative to radioprotection 
system, maybe useful even in wider considerations on risk.

Due to accepted linear-no-threshold  model for 
effects produced by ionizing radiation, 
justification of a practise has to be the first one posed, that is if 
the population exposure from military use of DU is justified or not. 
Comparison between dose estimates in such a scenario and dose limits and 
dose constraints stated by regulations is anyway useful, for a quantitative 
perception of risk.
In order to assess the need for remediation in contaminated areas, 
once again the question of justification has to be considered; 
{\it specific 
reference levels}, linked to the avertable annual dose, have to be defined 
by national authorities. "{\it Generic reference levels}
 ... should be used with 
great caution" and their use "should not prevent protective actions from 
being taken to reduce ... dominant components [of existing annual dose]" 
\cite{ICRP82}.
We next report a comment to assessed doses, comparing them with radiological 
and toxicological reference values, in order not to hold the question 
narrowed to exceeding of dose limits.

Values of annual dose in Tables \ref{tab1}, \ref{tab2} and \ref{tab3}
 show the same temporal shape, 
with an initial prevalent dose from irradiation by soil and a maximum 
from ingestion of contaminated drinking water occurring after about five 
hundreds years, when contamination reaches the acquifer serving the 
population. 
Maximum dose, progressively increasing as inventary of DU 
increases, is always lower than annual population limit (1 $mSv/y$), 
starts to be comparable with EPA cleanup limit criterion (150 $\mu Sv/y$,
\cite{EPA}) 
for 20 $cm$ depth. Exceeding of
dose constraint of $0.1~ mSv/y$ indicated in \cite{ICRP82} for
longlived isotopes may not be excluded.
This in general happen 
only after long times, due to the low mobility of the uranium oxides
(the mean transit times for insoluble uranium in the top 10
cm of soil range from 7.4  to 15.4 years with an average of 13.4
years \cite{27};  soluble forms have a mean transit times of one
month). 
At the maximum dose time concentration of DU in the 
water reaches the provisional value of WHO guideline for 
drinkable water (0.05 $Bq/l$ \cite {WHO}) already for 1 $cm$ depth. The 
concentration of DU in leafy vegetables at time of maximum 
dose ranges from 2 to 4 $Bq/kg$; no derived limit is defined for 
consumption of dietary parts.

Inhalation of highly contaminated soil may leave to exceeding 
of annual dose limit, with possible occurring of toxicological 
damage: maximum allowed concentration in air for workplaces stated 
by NRC, 45 $\mu g/m^3$ for soluble and 200 $\mu g/m^3$ for
 insoluble uranium 
forms, would be exceeded if dust loading was more than 1700 $\mu g/m^3$, 
a high but not extreme value. 
Less important seems ingestion of contaminated soil, due to the 
lower value of dose conversion factor with respect to the inhalation one. 
Anyway, ingestion of 1 $g$ maximum contaminated soil would result in 
120 $mg$ DU ingestion, when maximum daily ingestion of uranium, 
due to toxicological effects, was stated in 150 $mg$ by italian 
legislation till year 2000.

\section{Conclusions}

DU contained in projectiles, spread out in air after striking the target, 
falls out producing environmental and food chain contamination. 
Possible occurring of chemical hazard and entity of radiation dose must be 
assessed for different kind of exposure of people living in the area, 
taking into account both average and critical group exposure.
While waiting for measurements of contamination in Iraq, Bosnia, 
 Kosovo and Serbia, we have  computed radiation doses 
and concentrations for different contaminated soil thickness, 
as soil mixing will extend the initial superficial deposition to underlying 
layers in not undisturbed areas.

In order to assess the average exposure 
of population,
doses and concentrations have been obtained from an average 
value of soil contamination.
For the individuation of the critical group inhomogeneous soil
concentration has been considered.
The presented dose assessment   suggests
a  short term exposure due to inhalation and/or ingestion of contaminated 
soil and a long term exposure due to ingestion of contaminated water
and food;   the propagation of
the superficial contamination to the watertable critically depends
 on various hydrogeological parameters to be evaluated  on the site. 

In sites targeted by DU munitions special cautions
have to be adopted to reduce unaware exposures and cleanup 
must be planned on the basis of the measured concentrations.

\end{document}